\documentclass[aps,showpacs,preprintnumbers,amsmath, amssymb]{revtex4}
\usepackage{booktabs}

\usepackage{float}
\usepackage{graphics,epsfig}
\usepackage{graphicx}
\usepackage{dcolumn}
\usepackage{bm}

\begin{document}
\baselineskip=0.8 cm

\title{{\bf Bounds on the minimum orbital period of four-dimensional black holes: From singular to non-singular spacetimes}}

\author{Guohua Liu$^{1}$\footnote{Corresponding author: liuguohua1234@163.com}}
\author{Yan Peng$^{1,2}$\footnote{Corresponding author: yanpengphy@163.com}}
\author{Hua Huang$^{1}$}
\author{Fangyuan Yang$^{1}$}
\affiliation{\\$^{1}$ College of Mathematics and Physics, Xinjiang Agricultural University, Urumqi, Xinjiang, 830052,P.R.China}
\affiliation{\\$^{2}$ School of Mathematical Sciences, Qufu Normal University, Qufu, Shandong 273165,P.R.China}

\vspace*{0.2cm}
\begin{abstract}
\baselineskip=0.6 cm
\begin{center}
{\bf Abstract}
\end{center}

It was conjectured that the lower bound $T_{\min}\geq 4\pi M $ may be a universal
property of four-dimensional black holes, where $M$ is the black hole mass
and $T_{\min}$ is the minimum orbital period around black holes.
Recently, we established an upper bound $T_{\min}\le 6\sqrt{3}\pi M$
for Schwarzschild, Reissner--Nordstr\"om and Kerr--Newman black holes
\cite{YPL}. In this work we incorporate that analysis, which is not intended
for separate publication, into a unified treatment and extend it to
the non-singular Hayward and Bardeen black holes, demonstrating that
the combined bounds $4\pi M\le T_{\min}\le 6\sqrt{3}\pi M$ hold for
all five types of four-dimensional black holes investigated here.
For all types of studied black holes,
the upper bound is saturated by the standard Schwarzschild black hole and the
lower bound by the maximally rotating Kerr black hole, suggesting that these
bounds $4\pi M \leq T_{\min} \leq 6\sqrt{3}\pi M$
may be a universal property of four-dimensional black holes with horizons.

\end{abstract}

\maketitle
\newpage
\vspace*{0.2cm}

\section{Introduction}

Since the early days of general relativity, motion of test particles has been investigated in curved
black hole spacetimes \cite{NG1,NG2,NG3,NG4,NG5}. The study of test particle motion provides valuable
insights into understanding the nature of black holes and the surrounding spacetime structure
\cite{PR1}-\cite{PR14}. In recent years, significant progress has been made in the motion of
test particles around black holes. An interesting question is to study the minimum orbital
period around black holes and bounds on the minimum orbital period provide a new perspective
on the dynamics of particles in these extremely curved spacetimes.

Previous research has focused on establishing lower bounds on the minimum orbital
period of black holes.
For instance, Hod analytically studied the minimum orbital period as measured by observers
at asymptotic infinity around Schwarzschild and Kerr black holes
\cite{Hod1,Hod2}. For general rotating Kerr black holes, the situation is
complex due to the presence of angular momentum. However, Hod provided a rigorous analytical
proof that the orbital period $T(r)$ cannot be less than the lower bound $4\pi M$.
This bound is saturated by maximally rotating (extremal) Kerr black holes.
It was further conjectured that the
lower bound $4\pi M$ on minimum orbital periods may be a general property in highly curved
spacetimes \cite{Hod1}. In fact, it was found that this lower bound also exists for more
complex Kerr-Newman black holes \cite{Yan Peng}.
However, whether an upper bound exists for the minimum orbital period of
black holes has remained open until recently,
when we presented the first upper-bound
analysis of Schwarzschild, Reissner--Nordstr\"om and Kerr--Newman black
holes in \cite{YPL}.
Establishing such an upper bound
provides a clearer understanding of the possible range of minimum orbital
periods in black hole spacetimes. It is then natural to ask whether the
combined bounds $4\pi M\le T_{\min}\le 6\sqrt{3}\pi M$ hold more
generally, in particular for black holes without central singularities.

Unlike classical black holes, the Hayward and Bardeen black holes
represent special non-singular black hole models,
meaning they do not possess a central singularity \cite{HB1,HB2,BB}.
This property is achieved by introducing a specific form of matter field in
the Hayward black hole model and a magnetic charge parameter in the Bardeen black hole
model \cite{HB3,BB1,BB2,VV}. These modifications result in a
smoother spacetime geometry in the central regions of the black holes, thereby ensuring
that the spacetime curvature remains finite throughout their interiors. Such features
offer a more physically reasonable description of processes occurring within black holes.
Due to their non-singular nature, these solutions provide alternative perspectives for
investigating fundamental issues. For example, the thermodynamics of a modified Hayward
black hole under thermal fluctuations was investigated in \cite{RR1} and
particle acceleration processes in rotating versions of both Hayward and Bardeen black
holes were studied in \cite{RR2}. These two types of black
holes are of great research value. We intend to investigate whether the minimum orbital periods
are bounded in non-singular black hole spacetimes.

In this work we merge and extend these two lines of analysis into a
single treatment. The upper-bound results of \cite{YPL} are incorporated
together with the bounds for non-singular black holes, and we demonstrate
that the combined bounds $4\pi M\le T_{\min}\le 6\sqrt{3}\pi M$ hold for
all five types of four-dimensional black holes investigated here. We
summarize our main results at the end of the paper.

\section{Upper bound on the minimum orbital period of Schwarzschild black holes}

The Schwarzschild black hole is a non-rotating and uncharged black hole in general relativity.
We consider the static spherically symmetric Schwarzschild black hole in the asymptotically
flat background. The geometry of the Schwarzschild black hole is:
\begin{equation}
ds^{2}=-\left(1-\frac{2M}{r}\right)dt^{2}+\left(1-\frac{2M}{r}\right)^{-1}dr^{2}+r^{2}(d\theta^{2}+\sin^{2}\theta \, d\phi^{2}),
\label{eq:sch-metric}
\end{equation}
where $M$ is the black hole mass. And the event horizon is located at $r=2M$, which is determined by $1-\frac{2M}{r}=0$.
We study the physical and mathematical properties of the orbital period in the background of Schwarzschild black holes.
In the following, we would like to search for the circular trajectory with the shortest orbital
period as measured by asymptotic infinity observers.

In order to minimize the orbital period for a given radius $r$, one should move as
close as possible to the speed of light,
i.e., the spacetime interval $ds^{2}$ vanishes.
So the circular trajectory with the shortest
orbital period corresponds to $ds=dr=d\theta=0$. We consider test particles traveling
within the equatorial plane $\theta=\frac{\pi}{2}$. By restricting the motion of the
test particles to this plane, the original equation (\ref{eq:sch-metric}) can be transformed into
\begin{equation}
-\left(1-\frac{2M}{r}\right)dt^{2}+r^{2}d\phi^{2}=0.
\label{eq:sch-light}
\end{equation}

We take $d\phi=2\pi$ to describe one complete orbit. The orbital period $T(r)=dt$
describes the time for a particle to complete one full orbit as measured by
asymptotic infinity observers. It can be expressed by
\begin{equation}
T(r)=\frac{2\pi r}{\sqrt{1-\frac{2M}{r}}}.
\label{eq:sch-T}
\end{equation}

To find the minimum orbital period, we differentiate $T(r)$
with respect to the radius $r$ and set the derivative equal to zero:
\begin{equation}
T'(r)=\frac{d}{dr}\left(\frac{2\pi r}{\sqrt{1-\frac{2M}{r}}}\right)=\frac{2\pi(r-3M)}{(r-2M)\sqrt{1-\frac{2M}{r}}}=0.
\label{eq:sch-dT}
\end{equation}
This equation yields the critical radius:
\begin{equation}
r=3M,
\label{eq:sch-rc}
\end{equation}
indicating a local maximum or minimum value.

In order to determine whether $T(r)$ takes the minimum value at $r=3M$, we further calculate the second derivative $T''(r)$ at $r=3M$. The result is as follows:
\begin{equation}
T''(r)\big|_{r=3M}=\frac{6M^{2}\pi\sqrt{1-\frac{2M}{r}}}{(r-2M)^{3}}\bigg|_{r=3M}=\frac{2\sqrt{3}\pi}{M}>0.
\label{eq:sch-d2T}
\end{equation}
Therefore, at the point $r=3M$, the orbital period $T(r)$ takes the minimum value
\begin{equation}
T_{\min}=6\sqrt{3}\pi M.
\label{eq:sch-Tmin}
\end{equation}

\section{Upper bound on the minimum orbital period of Reissner-Nordstr\"{o}m black holes}

A Reissner-Nordstr\"{o}m black hole is a non-rotating, electrically charged black hole
described in general relativity. We consider the static spherically symmetric
Reissner-Nordstr\"{o}m black hole in the asymptotically flat background.
The line element of a charged Reissner-Nordstr\"{o}m black hole is:
\begin{equation}
ds^{2}=-\left(1-\frac{2M}{r}+\frac{Q^{2}}{r^{2}}\right)dt^{2}+\left(1-\frac{2M}{r}+\frac{Q^{2}}{r^{2}}\right)^{-1}dr^{2}+r^{2}(d\theta^{2}+\sin^{2}\theta \, d\phi^{2}),
\label{eq:rn-metric}
\end{equation}
where $M$ and $Q$ are respectively the mass and charge of the black hole.
The outer and inner horizons are located at $r_{h}=M\pm\sqrt{M^{2}-Q^{2}}$
obtained from $1-\frac{2M}{r}+\frac{Q^{2}}{r^{2}}=0$.

We consider the minimum orbital period of objects in the equatorial
plane of Reissner-Nordstr\"{o}m black holes. To simplify the problem,
we set $ds=dr=d\theta=0$ and $\theta=\frac{\pi}{2}$.
Then the equation (\ref{eq:rn-metric}) can be transformed into
\begin{equation}
-\left(1-\frac{2M}{r}+\frac{Q^{2}}{r^{2}}\right)dt^{2}+r^{2}d\phi^{2}=0.
\label{eq:rn-light}
\end{equation}

We take $d\phi=2\pi$ to describe a full circle orbit.
The orbital period $T(r)=dt$ as measured by asymptotic infinity observers is given by
\begin{equation}
T(r)=\frac{2\pi r}{\sqrt{1-\frac{2M}{r}+\frac{Q^{2}}{r^{2}}}}.
\label{eq:rn-T}
\end{equation}

To determine the minimum orbital period, we differentiate $T(r)$
with respect to $r$ and set the derivative equal to zero.
The derivative of $T(r)$ is
\begin{equation}
T'(r)=\frac{d}{dr}\left(\frac{2\pi r}{\sqrt{1-\frac{2M}{r}+\frac{Q^{2}}{r^{2}}}}\right)=\frac{2\pi r(r^{2}-3Mr+2Q^{2})}{(r^{2}-2Mr+Q^{2})^{3/2}}=0.
\label{eq:rn-dT}
\end{equation}

Solving $T'(r)=0$ gives the critical radii, which are
the candidate locations of the minimum orbital period.
The critical radii are
\begin{equation}
r=\frac{1}{2}\left(3M\pm\sqrt{9M^{2}-8Q^{2}}\right).
\label{eq:rn-rc}
\end{equation}

However, the solution $r=\frac{1}{2}(3M-\sqrt{9M^{2}-8Q^{2}})$
is discarded because it does not lie outside the outer event horizon.
In order to determine whether $T(r)$ attains the minimum value at $r=\frac{1}{2}(3M+\sqrt{9M^{2}-8Q^{2}})$,
we further calculate the second derivative of $T(r)$ as
\begin{equation}
T''(r)=\frac{2\pi(3M^{2}r^{2}-Q^{2}r^{2}-4MQ^{2}r+2Q^{4})}{(r^{2}-2Mr+Q^{2})^{5/2}},
\label{eq:rn-d2T}
\end{equation}
which helps in determining whether it is a minimum or maximum period.

Putting $r=\frac{1}{2}(3M+\sqrt{9M^{2}-8Q^{2}})$ into (\ref{eq:rn-d2T}), we find that $T''(r)>0$ for $0\leq Q\leq M$. So $T(r)$
takes the minimum value at $r=\frac{1}{2}(3M+\sqrt{9M^{2}-8Q^{2}})$. At this point, the minimum value of $T(r)$ is
\begin{equation}
T_{\min}=\frac{\pi(3M+\sqrt{9M^{2}-8Q^{2}})^{2}}{\sqrt{6M^{2}-4Q^{2}+2M\sqrt{9M^{2}-8Q^{2}}}}.
\label{eq:rn-Tmin}
\end{equation}

In order to get the upper bound of $T_{\min}$, we calculate $\frac{T_{\min}}{M}$ as a
function of $\frac{Q}{M}$. We find that $\frac{T_{\min}}{M}$ decreases as we increase
the value of $\frac{Q}{M}$. In the case of $\frac{Q}{M}=0$, the upper bound is $32.6484$,
which coincides exactly with the Schwarzschild value
$6\sqrt{3}\pi$ since $6\sqrt{3}\pi\approx 32.6484$. Our results imply
that there is an upper bound on the minimum orbital period of Reissner-Nordstr\"{o}m black holes in the form
\begin{equation}
T_{\min}\leq 6\sqrt{3}\pi M.
\label{eq:rn-bound}
\end{equation}

\begin{figure}[htbp]
\centering
\includegraphics[scale=0.7]{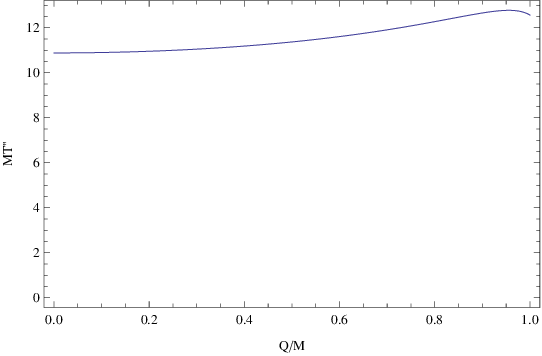}
\includegraphics[scale=0.7]{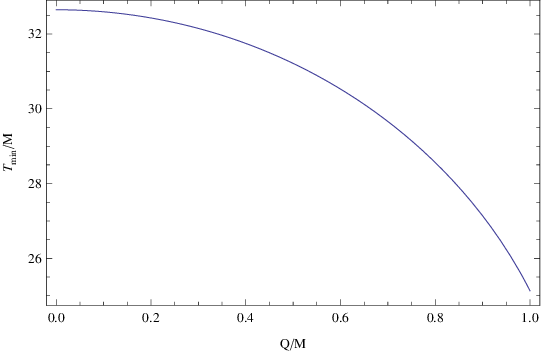}
\caption{Left panel: $MT''$ as a function of $Q/M$ for Reissner-Nordstr\"{o}m black holes,
confirming $T''(r)>0$ in the physical range $0\leq Q\leq M$. Right panel: $T_{\min}/M$ as
a function of $Q/M$, showing that the normalized minimum orbital period decreases with
increasing charge and approaches the Schwarzschild upper bound $6\sqrt{3}\pi$ at $Q=0$. }
\label{fig:RN}
\end{figure}

The decrease of $T_{\min}$ with $Q$ is analytic rather than merely
numerical. A direct differentiation of Eq.~(14) with respect to $Q$ is
hindered by nested square roots, since $Q$ appears both inside and
outside $\sqrt{9M^{2}-8Q^{2}}$. For simplicity, we introduce the
one-to-one change of variable
\begin{equation}
u=\sqrt{9M^{2}-8Q^{2}},\qquad
u\in[\,M,\,3M\,]\ \ \text{as}\ \ Q\in[0,M],
\end{equation}
which is strictly decreasing in $Q$ ($du/dQ=-8Q/u<0$).
As $Q^{2}=(9M^{2}-u^{2})/8$,
the denominator of Eq.~(14) factorizes as
\begin{equation}
6M^{2}-4Q^{2}+2Mu=\frac{(u+M)(u+3M)}{2},
\end{equation}
and Eq.~(14) takes the form
\begin{equation}
T_{\min}(u)=\sqrt{2}\,\pi\,(3M+u)^{3/2}\,(u+M)^{-1/2}.
\end{equation}

we first take the logarithm of $T_{\min}$. This does not change the
monotonicity as the logarithm is a strictly increasing function.
Since
\begin{equation}
\frac{d\ln T_{\min}}{du}
 =\frac{3}{2(3M+u)}-\frac{1}{2(u+M)}
 =\frac{u}{(3M+u)(u+M)}>0 ,
\end{equation}
$T_{\min}$ increases strictly with $u$, while $u$ decreases strictly
with $Q$ ($du/dQ=-8Q/u<0$). So we find that
\begin{equation}
\frac{d T_{\min}}{d Q}<0 \qquad (0<Q\le M),
\end{equation}
i.e.\ $T_{\min}$ is a strictly decreasing function of the charge.
So we obtain the bounds
\begin{equation}
T_{\min}(0)=6\sqrt{3}\pi M,\qquad
T_{\min}(M)=8\pi M,
\end{equation}
which reproduce the Schwarzschild upper bound and fix the lower endpoint
$8\pi$ quoted in Table~I.

\section{Upper bound on the minimum orbital period of Kerr-Newman black holes}

The Kerr-Newman black hole is a type of black hole that possesses both rotation
and electric charge. It is characterized by three key parameters: the black hole mass $M$,
its electric charge $Q$, and its angular momentum per unit
mass $a$. In the Boyer-Lindquist coordinate system, the metric for a
Kerr-Newman black hole is given by the following expression \cite{NG2}:
\begin{equation}
\begin{split}
ds^{2}=&-\left(1-\frac{2Mr-Q^{2}}{\rho^{2}}\right)dt^{2}-\frac{2a(2Mr-Q^{2})\sin^{2}\theta}{\rho^{2}}dt \, d\phi \\
&+\frac{\rho^{2}}{\Delta}dr^{2}+\rho^{2}d\theta^{2}+\left[r^{2}+a^{2}+\frac{a^{2}(2Mr-Q^{2})\sin^{2}\theta}{\rho^{2}}\right]\sin^{2}\theta \, d\phi^{2},
\end{split}
\label{eq:kn-metric}
\end{equation}
where $\rho^{2}=r^{2}+a^{2}\cos^{2}\theta$ and $\Delta=r^{2}-2Mr+a^{2}+Q^{2}$.
The location of the black hole event horizon is determined by the condition
$\Delta=r^{2}-2Mr+a^{2}+Q^{2}=0$. The outer and inner event horizon radii
of a Kerr-Newman black hole are calculated using the formula
$r_{\text{horizon}}=M\pm\sqrt{M^{2}-a^{2}-Q^{2}}$.

On the equatorial plane with $\theta=\frac{\pi}{2}$, the functions
can be simplified as $\sin\theta=1$, $\cos\theta=0$ and $\rho^{2}=r^{2}$.
For objects traveling in circular paths around such black holes,
the circular orbit condition is $dr=d\theta=0$. Under these conditions,
the metric of the Kerr-Newman black hole can be simplified significantly and transformed into
\begin{equation}
ds^{2}=-\left(1-\frac{2Mr-Q^{2}}{r^{2}}\right)dt^{2}-\frac{2a(2Mr-Q^{2})}{r^{2}}dt \, d\phi+\left[r^{2}+a^{2}+\frac{a^{2}(2Mr-Q^{2})}{r^{2}}\right]d\phi^{2}.
\label{eq:kn-eq}
\end{equation}

At the light speed limit, there is $ds^{2}=0$, and the relation (\ref{eq:kn-eq}) becomes:
\begin{equation}
-\left(1-\frac{2Mr-Q^{2}}{r^{2}}\right)dt^{2}-\frac{2a(2Mr-Q^{2})}{r^{2}}dt \, d\phi+\left[r^{2}+a^{2}+\frac{a^{2}(2Mr-Q^{2})}{r^{2}}\right]d\phi^{2}=0.
\label{eq:kn-light}
\end{equation}

To analyze the orbital period of a circular orbit in the equatorial plane of a rotating,
charged black hole, we set $d\phi=2\pi$ to represent a full orbital period.
The time period $T(r)=dt$ is derived from the following equation:
\begin{equation}
-\left(1-\frac{2Mr-Q^{2}}{r^{2}}\right)T(r)^{2}-\frac{4\pi a(2Mr-Q^{2})}{r^{2}}T(r)+4\pi^{2}\left[r^{2}+a^{2}+\frac{a^{2}(2Mr-Q^{2})}{r^{2}}\right]=0.
\label{eq:kn-quad}
\end{equation}

Solving this equation for $T(r)$, we obtain the expression for the orbital period:
\begin{equation}
T(r)=2\pi\frac{\sqrt{r^{2}-2Mr+a^{2}+Q^{2}}-\frac{2Ma}{r}+\frac{aQ^{2}}{r^{2}}}{1-\frac{2M}{r}+\frac{Q^{2}}{r^{2}}}.
\label{eq:kn-T}
\end{equation}
Of the two algebraic branches, we take the branch, which gives
the shorter period for $a\geq0$.
This result highlights that the orbital period depends on several key parameters:
the radius $r$, black hole mass $M$, charge $Q$ and angular momentum $a$.

To find the minimum value of $T_{min}$, we calculate the derivative of the orbital
period with respect to the radial coordinate $r$. The derivative is derived as:
\begin{equation}
\frac{dT}{dr} = \frac{2\pi r \left[ 2Q^4 + Q^2r(3r-7M ) + 2a^2(Q^2 - Mr) + r^2(6M^2 - 5Mr + r^2) - 2a(Q^2 - Mr)\sqrt{a^2 + Q^2 - 2Mr + r^2} \right]}{(Q^2 + r(r-2M ))^2 \sqrt{a^2 + Q^2 + r(r-2M)}}.
\end{equation}
This derivative provides insight into how the orbital period changes with $r$.
Setting the derivative to zero, we can identify the critical points where the orbital
period reaches its minimum value. However, due to the equation's complexity,
numerical methods are needed. With various values of normalized charge $\frac{Q}{M}$
and the normalized angular momentum $\frac{a}{M}$, we find that $\frac{T_{\min}}{M}$ decreases as
we increase $\frac{Q}{M}$ or $\frac{a}{M}$.
The upper bound on the minimum orbital period is therefore achieved in the Schwarzschild limit.

We plot the normalized minimum orbital period $T_{\min}/M$ as a function of the
normalized charge $Q/M$ and normalized angular momentum $a/M$. As can be seen from Fig. 2,
the upper bound is obtained at $Q/M=0$ and $a/M=0$. The calculated
numerical upper bound $32.6484$ coincides exactly with the upper bound $6\sqrt{3}\pi$ (since $6\sqrt{3}\pi\approx 32.6484$).
Our results again suggest that there may be a general upper bound on minimum orbital period for black hole spacetimes in the form
\begin{equation}
T_{\min}\leq 6\sqrt{3}\pi M.
\label{eq:kn-bound}
\end{equation}

\begin{figure}
    \centering 
    \includegraphics[scale=1.0]{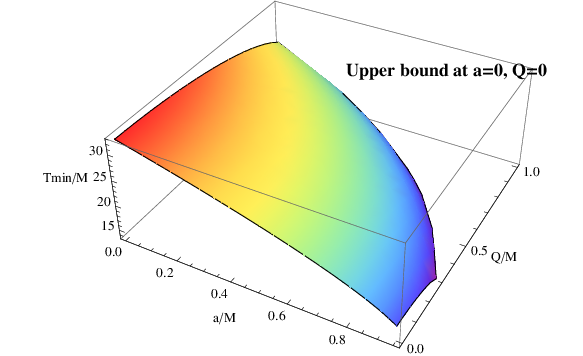} 
    \includegraphics[scale=0.65]{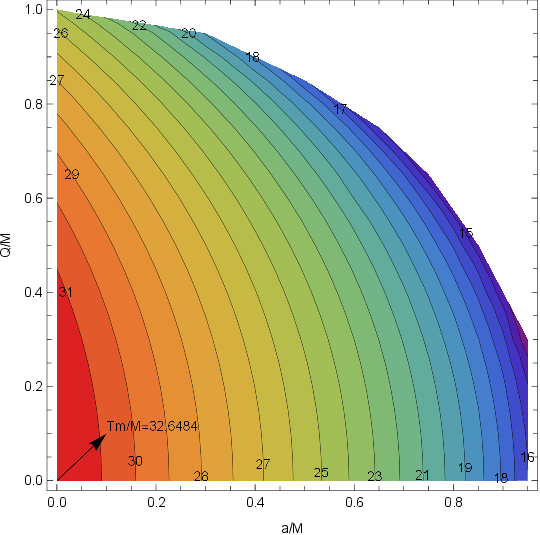}
    \caption{The left panel shows $\frac{T_{min}}{M}$ as a function of $\frac{Q}{M}$ and $\frac{a}{M}$.
    Each black curve in the right panel is a contour of constant $\frac{T_{min}}{M}$.
    The numerical upper bound $\frac{T_{min}}{M}=32.6484\thickapprox 6\sqrt{3}\pi$ is obtained in the case of $\frac{Q}{M}=0$ and $\frac{a}{M}=0$.
    The plotted domain is restricted to the physical region $a^2+Q^2\leq M^2$.
Outside this region, the spacetime contains a naked singularity rather than a black hole.} 
\end{figure}

The upper bound can in fact be established analytically, without
locating the position of the minimum. Since the outer horizon is
$r_{outer}=M+\sqrt{M^{2}-a^{2}-Q^{2}}\le 2M<3M$, the point $r=3M$ lies
outside the horizon for all admissible parameters, and therefore
$T_{\min}\le T(3M)$: the period evaluated at the single point $r=3M$
already provides an upper bound on $T_{\min}$ for every Kerr-Newman
black hole. Evaluating Eq.~(26) at $r=3M$ and simply setting $\alpha=a/M$,
$q=Q/M$ gives
\begin{equation}
T(3M)=\frac{18\pi M}{3+q^{2}}
\left[\sqrt{3+\alpha^{2}+q^{2}}-\frac{2\alpha}{3}
+\frac{\alpha q^{2}}{9}\right] .
\end{equation}
In order to judge the sign of $T(3M)-6\sqrt{3}\pi M$ directly, we
would have to deal with a fraction whose numerator and denominator both
depend on $\alpha$ and $q$. Since the denominator $3+q^{2}$ is always
positive, it is more convenient to move everything to one side and
define
\begin{equation}
G(\alpha,q)=3\sqrt{3+\alpha^{2}+q^{2}}-2\alpha+\frac{\alpha q^{2}}{3}
-\sqrt{3}\,(3+q^{2}) .
\end{equation}
Then $T(3M)\le 6\sqrt{3}\pi M$ is equivalent to $G\le 0$, so proving
the upper bound becomes finding the maximum of $G$ in the physical
region $\alpha\ge 0$, $\alpha^{2}+q^{2}\le 1$:
\begin{equation}
\frac{\partial G}{\partial\alpha}
=\frac{3\alpha}{\sqrt{3+\alpha^{2}+q^{2}}}-2+\frac{q^{2}}{3}
\le\frac{3}{2}-2+\frac{1}{3}=-\frac{1}{6}<0 ,
\end{equation}
where we have used $\alpha/\sqrt{3+\alpha^{2}}\le 1/2$ for
$0\le\alpha\le 1$ in the first step. In the same way, for $q>0$,
\begin{equation}
\frac{\partial G}{\partial q}
=q\left[\frac{3}{\sqrt{3+\alpha^{2}+q^{2}}}
+\frac{2\alpha}{3}-2\sqrt{3}\right]
\le q\left(\frac{2}{3}-\sqrt{3}\right)<0,
\end{equation}
where we have used $3/\sqrt{3+\alpha^{2}+q^{2}}\le\sqrt{3}$ and
$2\alpha/3\le 2/3$.

Thus $G$ decreases strictly along both $\alpha$ and $q$, and since
$G(0,0)=0$ it follows that $G\le 0$ throughout the physical region,
with equality only for $\alpha=q=0$. Consequently,
\begin{equation}
T_{\min}\le T(3M)\le 6\sqrt{3}\pi M .
\end{equation}
The minimum of the
Schwarzschild period was obtained analytically at $r=3M$, where
$T_{\min}=6\sqrt{3}\pi M$. Hence no universal constant smaller than
$6\sqrt{3}\pi M$ can bound $T_{\min}$, and the value $6\sqrt{3}\pi M$
is attained precisely in the non-rotating, uncharged limit. In fact,
if a black hole with $(\alpha,q)\neq(0,0)$ saturated the bound, then
$T_{\min}=6\sqrt{3}\pi M$ together with $T_{\min}\le T(3M)\le
6\sqrt{3}\pi M$ would force $T(3M)=6\sqrt{3}\pi M$ and therefore
$\alpha=q=0$, a contradiction. The upper bound quoted in Table~I is
thus exact rather than numerical: it is the universal upper
bound for the Kerr--Newman family, saturated by the
Schwarzschild black hole.

\section{Bounds on the minimum orbital period of Hayward black holes}

The Hayward black hole is a spherically symmetric non-singular solution
in general relativity. It can be viewed as a regularization of the Schwarzschild black hole,
incorporating quantum gravity effects through an additional parameter.
The metric of the Hayward black hole is expressed as \cite{HB1,HB2}:
\begin{equation}
ds^{2}=-\left(1-\frac{2Mr^{2}}{r^{3}+2L^{2}M}\right)dt^{2}+\left(1-\frac{2Mr^{2}}{r^{3}+2L^{2}M}\right)^{-1}dr^{2}+r^{2}(d\theta^{2}+\sin^{2}\theta \, d\phi^{2}),
\label{eq:hay-metric}
\end{equation}
where $M$ is the black hole mass and $L$ is the length parameter related to quantum gravity effects.
As $r\to 0$, the metric function $f(r)\to 1-r^{2}/L^{2}$ exhibits de Sitter-like behavior, avoiding a central
singularity. When $r\gg L$, it reduces to the Schwarzschild metric with
classical behaviors on large scales. This model is important for exploring the potential
effects of quantum gravity within the framework of general relativity.

In the following analysis, we obtain the period of circular orbits around the
Hayward black holes. For objects moving along circular paths around such black holes,
the condition for a circular orbit is $dr=d\theta=0$. Without loss of generality,
we assume that test objects travel within the equatorial plane $\theta=\frac{\pi}{2}$.
Under these conditions, the metric can be simplified to:
\begin{equation}
ds^{2}=-\left(1-\frac{2Mr^{2}}{r^{3}+2L^{2}M}\right)dt^{2}+r^{2}d\phi^{2}.
\label{eq:hay-eq}
\end{equation}

We aim to determine the minimum period of objects orbiting the black hole.
To explore this, we focus on objects approaching the speed of light. In the
context of general relativity, the light speed condition implies that the
spacetime interval $ds^{2}$ is zero. Under this condition, we obtain the
following relation:
\begin{equation}
-\left(1-\frac{2Mr^{2}}{r^{3}+2L^{2}M}\right)dt^{2}+r^{2}d\phi^{2}=0.
\label{eq:hay-light}
\end{equation}

Without loss of generality, we assume that the coordinate $\phi$ increases along the
direction of motion since the metric is spherically symmetric. The future-directed
condition requires $dt>0$ outside the outer horizon. With $d\phi>0$ and $dt>0$,
we obtain $dt/d\phi>0$ for the region outside the black hole.
Then equation (\ref{eq:hay-light}) leads to the relation:
\begin{equation}
dt=\frac{r}{\sqrt{1-\frac{2Mr^{2}}{r^{3}+2L^{2}M}}}d\phi.
\label{eq:hay-dt}
\end{equation}

To represent one full orbit around the black hole, we assume that the angle
coordinate $\phi$ increases from $0$ to $2\pi$. It allows us to analyze the
periodic motion of objects. Additionally, we denote the time period observed
from infinity as $T$. According to the equation (\ref{eq:hay-dt}),
the orbital period $T$ can be obtained from an integral:
\begin{equation}
\int_{0}^{T}dt=\int_{0}^{2\pi}\frac{r}{\sqrt{1-\frac{2Mr^{2}}{r^{3}+2L^{2}M}}}d\phi.
\label{eq:hay-int}
\end{equation}

Solving equation (\ref{eq:hay-int}), we obtain the orbital period:
\begin{equation}
T=\frac{2\pi r}{\sqrt{1-\frac{2Mr^{2}}{r^{3}+2L^{2}M}}}.
\label{eq:hay-T}
\end{equation}

In the case of $L=0$, the Hayward black hole metric reduces to the Schwarzschild
black hole case with the minimum orbital period $T_{\min}=6\sqrt{3}\pi M$.
In the case of $L\neq 0$, we determine the minimum value $T_{\min}$ numerically.
The derivative is important in understanding how the orbital period varies
with $r$. The derivative is calculated as follows:
\begin{equation}
\frac{dT}{dr}=\frac{2\pi(r^{6}-3Mr^{5}+4L^{2}Mr^{3}+4L^{4}M^{2})\sqrt{1-\frac{2Mr^{2}}{r^{3}+2L^{2}M}}}{(r^{3}-2Mr^{2}+2L^{2}M)^{2}}.
\label{eq:hay-dT}
\end{equation}

By setting the derivative to be zero, we can identify the critical points where the orbital period
reaches its minimum value. So we set $\frac{dT}{dr}=0$ and solve it for the region outside the outer horizon.
However, due to the equation's complexity, numerical methods are needed to find a solution.
Through extensive numerical analysis with various values of $M$ and $L$, it is observed that $T_{\min}/M$ decreases as $L$ increases.
So the upper bound on minimum orbital period should be achieved in the Schwarzschild limit.

The monotonic decrease of $T_{\min}$ with $L$ can also be established
analytically for $L\ll M$. The metric function depends on $L^{2}$ and it can be expanded as
\begin{equation}
f(r)=1-\frac{2M r^{2}}{r^{3}+2L^{2}M}
    =1-\frac{2M}{r}+\frac{4L^{2}M^{2}}{r^{4}}+O(L^{4}) .
\end{equation}

The minimum $T_{\min}(L)=T(r_{*}(L),L)$ depends on $L$ through two
channels: the direct dependence of $T$ on $L$, and the shift of the
extremal radius $r_{*}(L)$. In order to differentiate it, we first
write down the chain rule
\begin{equation}
\frac{d T_{\min}}{d (L^{2})}
=\frac{\partial T}{\partial r}\bigg|_{r_{*}}
 \frac{d r_{*}}{d (L^{2})}
+\frac{\partial T}{\partial (L^{2})}\bigg|_{r_{*}} .
\end{equation}
Since $r_{*}$ is the minimum point, we have
$\partial T/\partial r|_{r_{*}}=0$.
Considering $T=2\pi r f^{-1/2}$ from eq.~(39) together with $\partial T/\partial r|_{r_{*}}=0$,
the first term vanishes and
relation (42) becomes:
\begin{equation}
\frac{d T_{\min}}{d (L^{2})}
=\frac{\partial T}{\partial (L^{2})}\bigg|_{r_{*}}
=-\pi r_{*}\, f(r_{*})^{-3/2}\,
\frac{\partial f}{\partial (L^{2})}\bigg|_{r_{*}}
=-\frac{4\pi M^{2}}{r_{*}^{3}\, f(r_{*})^{3/2}} .
\end{equation}

For $L=0$, the minimum occurs at $r_{*}=3M$ with $f(3M)=1/3$ and $T_{\min}(0)=6\sqrt{3}\pi M$.
Substituting these values into eq.~(43), we find
$dT_{\min}/d(L^{2})|_{L=0}=-4\sqrt{3}\pi/(9M)$. Since $T_{\min}$
depends on $L^{2}$, we can expand it around $L=0$ as
\begin{equation}
T_{\min}(L)=6\sqrt{3}\pi M
           -\frac{4\sqrt{3}\pi}{9}\,\frac{L^{2}}{M}+O(L^{4}) .
\end{equation}
The leading correction is strictly negative, hence the upper bound
$6\sqrt{3}\pi M$ is approached from below as $L\to0$, in agreement with
the numerical results in Fig.~3.

We need to examine whether large $L$ can violate the lower bound $4\pi M$. If $L$ becomes excessively large,
the horizon disappears, and the configuration can no longer be classified as a black hole. Therefore,
it is essential to calculate the critical value of $L$ that supports the existence of a black hole horizon.

From the metric function $f(r)=1-\frac{2Mr^{2}}{r^{3}+2L^{2}M}$, the horizon can be determined by $f(r)=0$, which gives $r^{3}-2Mr^{2}+2L^{2}M=0$. To find the critical $L$ above which no horizon exists, we require this cubic equation to have a degenerate root, which can be expressed as:
\begin{equation}
h_{1}(r)=r^{3}-2Mr^{2}+2L^{2}M=0, \qquad h_{1}'(r)=3r^{2}-4Mr=0\Rightarrow r=\frac{4M}{3}.
\label{eq:hay-crit}
\end{equation}

Substituting $r=\frac{4M}{3}$ into $h_{1}(r)=0$, we obtain the relations:
\begin{equation}
\left(\frac{4M}{3}\right)^{3}-2M\left(\frac{4M}{3}\right)^{2}+2L^{2}M=0 \quad \Rightarrow \quad L_{\text{critical}}=\frac{4M}{3\sqrt{3}}\approx 0.7698M.
\label{eq:hay-Lcrit}
\end{equation}
We have numerically confirmed that no horizon exists when $L$ is above the critical value $L_{\text{critical}}$.

Then, we search for the minimum orbital period outside the black hole horizons.
We represent the normalized minimum orbital period $T_{\min}/M$ as a function
of $M$ and $L$ in Fig.3. The smallest value of the normalized minimum orbital
period is found to be $30.89$, which is well above Hod's lower bound $4\pi\approx 12.56$ \cite{Hod1,Hod2,Yan Peng}.
The upper bound is obtained at $L=0$, where the calculated numerical upper bound is $32.6484$,
nearly identical to $6\sqrt{3}\pi$ (since $6\sqrt{3}\pi\approx 32.6484$). In fact, this upper
bound has also been analytically obtained from the Schwarzschild case. These results reinforce
the conjecture that there may be general bounds on the minimum orbital period expressed as
\begin{equation}
4\pi M \leq T_{\min} \leq 6\sqrt{3}\pi M.
\label{eq:hay-bound}
\end{equation}

\begin{figure}[htbp]
\centering
\includegraphics[scale=0.80]{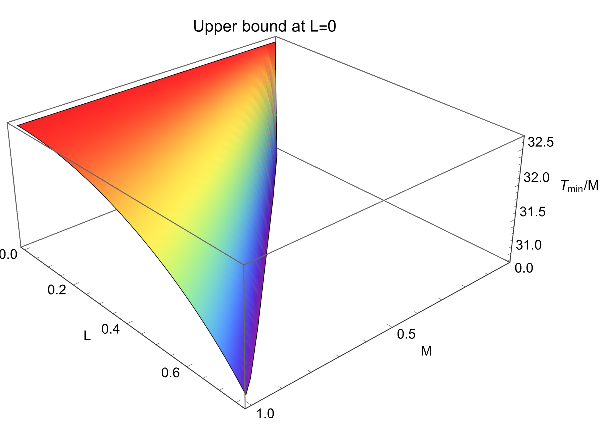}
\includegraphics[scale=0.65]{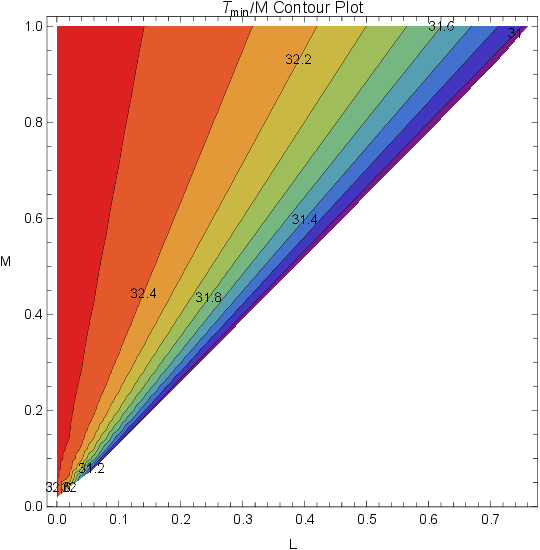}
\caption{Left panel: $T_{\min}/M$ as a function of $M$ and $L$ for Hayward black holes. Right panel: contour plot of $T_{\min}/M$.
The numerical upper bound $T_{\min}/M=32.6484\approx 6\sqrt{3}\pi$ is obtained in the case of $L=0$.}
\label{fig:Hayward}
\end{figure}

\section{Bounds on the minimum orbital period of Bardeen black holes}

The Bardeen black hole is another well-known non-singular black hole model
within general relativity. Unlike classical black holes, the Bardeen black hole
spacetime remains smooth even at the center. This regularization is achieved by
introducing a magnetic monopole charge that modifies the mass function in a way
consistent with the weak energy condition. The Bardeen black hole
can be described by \cite{BB}:
\begin{equation}
ds^{2}=-\left(1-\frac{2m}{r}\right)dt^{2}+\left(1-\frac{2m}{r}\right)^{-1}dr^{2}+r^{2}(d\theta^{2}+\sin^{2}\theta \, d\phi^{2}),
\label{eq:bar-metric}
\end{equation}
where $m=m(r)=M\left(\frac{r^{2}}{r^{2}+g^{2}}\right)^{3/2}$. Here $M$
is the total ADM mass of the black hole and $g$ denotes a parameter related
to the magnetic charge associated with the nonlinear electrodynamics field.
As $g\to 0$, the solution reduces to the standard Schwarzschild solution.

Without loss of generality, we consider equatorial circular orbits around black holes.
In such cases, the conditions $dr=d\theta=0$ and $\theta=\frac{\pi}{2}$ hold.
The spacetime metric simplifies to be
\begin{equation}
ds^{2}=-\left(1-\frac{2m}{r}\right)dt^{2}+r^{2}d\phi^{2}.
\label{eq:bar-eq}
\end{equation}

We are interested in the shortest orbital period for objects moving near
the speed of light satisfying $ds^{2}=0$. Applying this light speed limit
condition, we obtain the relation
\begin{equation}
-\left(1-\frac{2m}{r}\right)dt^{2}+r^{2}d\phi^{2}=0.
\label{eq:bar-light}
\end{equation}

To find the orbital period $T$, we also use an integral approach.
We integrate over a full orbit from $\phi=0$ to $\phi=2\pi$.
Therefore, the orbital period can be obtained as an integral:
\begin{equation}
\int_{0}^{T}dt=\int_{0}^{2\pi}\frac{r}{\sqrt{1-\frac{2m}{r}}}d\phi.
\label{eq:bar-int}
\end{equation}

From this equation, we obtain the expression for the orbital period:
\begin{equation}
T=\frac{2\pi r}{\sqrt{1-\frac{2m}{r}}},
\label{eq:bar-T}
\end{equation}
where $m(r)=M\left(\frac{r^{2}}{r^{2}+g^{2}}\right)^{3/2}$. This expression
shows the relationship between the orbital period and parameters such as the
radial distance $r$, the black hole mass $M$ and the magnetic charge parameter $g$.

To search for the minimum value $T_{\min}$, we calculate the derivative of
the orbital period with respect to $r$. Setting this derivative to be zero
helps us find the possible critical points where the orbital period reaches
its minimum value. Since the equation is too complicated,
we resort to numerical methods.
Our results show that $T_{\min}/M$ decreases as the
parameter $g$ increases.
This means that the upper bound for the minimum
orbital period is achieved in the case of Schwarzschild black holes with
$g=0$. If $g$ becomes too large, the black hole's horizon disappears,
and the configuration is no longer a black hole.

The same argument applies to small $g$ as well. In order to expand
the mass function, we first rewrite it as
$\left(1+g^{2}/r^{2}\right)^{-3/2}$ by dividing the numerator and the
denominator by $r^{2}$, and then use the binomial expansion
$(1+x)^{-3/2}=1-\frac{3}{2}x+O(x^{2})$ with $x=g^{2}/r^{2}$. This
gives
\begin{equation}
m(r)=M\left(\frac{r^{2}}{r^{2}+g^{2}}\right)^{3/2}
=M\left(1+\frac{g^{2}}{r^{2}}\right)^{-3/2}
=M-\frac{3Mg^{2}}{2r^{2}}+O(g^{4}).
\end{equation}
Substituting eq.~(53) into $f(r)=1-2m(r)/r$, the leading term $M$
gives the Schwarzschild part $1-2M/r$, while the $g^{2}$-order term
contributes a positive correction $3Mg^{2}/r^{3}$, so that
\begin{equation}
f(r)=1-\frac{2m(r)}{r}
    =1-\frac{2M}{r}+\frac{3Mg^{2}}{r^{3}}+O(g^{4}) .
\end{equation}
The same method applied in the Hayward case yields
\begin{equation}
\frac{d T_{\min}}{d (g^{2})}
 =-\pi r_{*}\, f(r_{*})^{-3/2}\,\frac{3M}{r_{*}^{3}},
\end{equation}
and at $r_{*}=3M$, it gives
\begin{equation}
T_{\min}(g)=6\sqrt{3}\pi M
           -\sqrt{3}\pi\,\frac{g^{2}}{M}+O(g^{4}) .
\end{equation}
The leading correction is again strictly negative, so the minimum orbital
period starts to decrease as soon as the magnetic charge parameter is
switched on.

For the Bardeen black hole, the horizon is determined by
$1-\frac{2m(r)}{r}=0$, where $m(r)=\frac{Mr^{3}}{(r^{2}+g^{2})^{3/2}}$. The horizon condition reduces to:
\begin{equation}
h_{2}(r)=r-\frac{2Mr^{3}}{(r^{2}+g^{2})^{3/2}}=0.
\label{eq:bar-horizon}
\end{equation}

In order to determine the critical conditions for the existence of a black hole horizon, we impose the necessary mathematical constraints:
\begin{equation}
h_{2}(r)=r-\frac{2Mr^{3}}{(r^{2}+g^{2})^{3/2}}=0, \qquad h_{2}'(r)=1-\frac{6Mr^{2}g^{2}}{(r^{2}+g^{2})^{5/2}}=0.
\label{eq:bar-crit}
\end{equation}

From these equations, we get the relations:
\begin{equation}
(r^{2}+g^{2})^{3/2}=2Mr^{2}, \qquad (r^{2}+g^{2})^{5/2}=6Mr^{2}g^{2}.
\label{eq:bar-crit1}
\end{equation}

These equations allow us to obtain:
\begin{equation}
6Mr^{2}g^{2}=(r^{2}+g^{2})^{5/2}=(r^{2}+g^{2})\cdot(r^{2}+g^{2})^{3/2}=(r^{2}+g^{2})\cdot 2Mr^{2}.
\label{eq:bar-comb}
\end{equation}

From this relation, we get $r^{2}=2g^{2}$. Putting this into the first equation, we find the critical value:
\begin{equation}
g_{\text{critical}}=\frac{4M}{3\sqrt{3}}\approx 0.7698M.
\label{eq:bar-gcrit}
\end{equation}
We have numerically found that there
is no horizon when $g$ is above the critical value $g_{\text{critical}}$.

It is worth commenting on the fact that the two critical values coincide,
$L_{\mathrm{critical}} = g_{\mathrm{critical}} = 4M/(3\sqrt{3})$,
although the Hayward and Bardeen metrics are quite different. This
coincidence can be traced to a common algebraic structure of the two
horizon equations. For the Hayward black hole, the horizon condition (45)
is a cubic polynomial in $r$ with a vanishing linear term,
$r^{3}-2Mr^{2}+2L^{2}M=0$. For the Bardeen black hole, squaring the
horizon condition (57) and introducing $w=r^{2/3}$, one finds
$w^{3}-(4M^{2})^{1/3}w^{2}+g^{2}=0$, which is again a cubic polynomial
with a vanishing linear term. For such a cubic with a vanishing linear term equation $X^{3}+aX^{2}+c=0$,
the degenerate-root condition is $4a^{3}+27c=0$, which in both cases
yields the same ratio and hence the same critical parameter,
$L_{\mathrm{critical}}^{2} = g_{\mathrm{critical}}^{2} = 16M^{2}/27$.
This indicates that the equality of the critical values reflects the
algebraic form of the two metrics.

After finding this critical value of $g$, we can then determine the minimum orbital period.
The results indicate that the minimum value of the normalized minimum orbital period is
approximately $28.43$. This value is higher than the lower bound proposed by Hod, which
is $4\pi\approx 12.56$ \cite{Hod1,Hod2,Yan Peng}. The upper bound is observed at $g=0$, with a
calculated value of approximately $32.6484$. This is very close to $6\sqrt{3}\pi$,
which is roughly $32.6484$. This upper limit is also derived
analytically from the Schwarzschild case. These results provide
bounds on the minimum orbital period for black hole spacetimes expressed as:
\begin{equation}
4\pi M \leq T_{\min} \leq 6\sqrt{3}\pi M.
\label{eq:bar-bound}
\end{equation}
In summary, the bounds hold for different
black hole models, including Schwarzschild, Reissner-Nordstr\"{o}m, Kerr-Newman (and hence Kerr),
non-singular Hayward and Bardeen black holes, indicating that these bounds might
be a universal feature of black hole spacetimes.

\begin{figure}[htbp]
\centering
\includegraphics[scale=0.80]{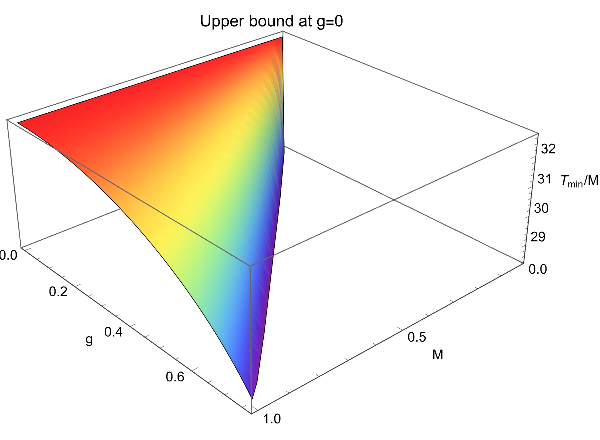}
\includegraphics[scale=0.65]{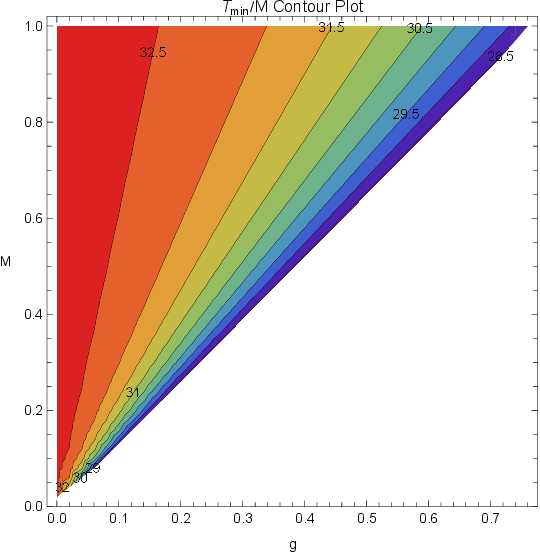}
\caption{Left panel: $T_{\min}/M$ as a function of $M$ and $g$ for Bardeen black holes. Right panel: contour plot of $T_{\min}/M$. The numerical upper bound $T_{\min}/M=32.6484\approx 6\sqrt{3}\pi$ is obtained in the case of $g=0$. }
\label{fig:Bardeen}
\end{figure}

\begin{table}[htbp]
\centering
\caption{Summary of the bounds on the minimum orbital period $T_{\min}/M$
for various four-dimensional black holes. The upper bound $6\sqrt{3}\pi$
is saturated in the Schwarzschild limit, while the lower bound $4\pi$
is saturated by the maximally rotating (extremal) Kerr black hole.}
\label{tab:summary}
\renewcommand{\arraystretch}{1.3}
\begin{tabular}{|l|l|l|l|}
\hline
Black hole & Parameter range & Range of $T_{\min}/M$ & Saturated bound \\
\hline
Schwarzschild &~~~~~~~~ --- & $T_{\min}/M = 6\sqrt{3}\pi$ & Upper bound \\
\hline
Reissner--Nordstr\"om & $0 \leq Q \leq M$ & $8\pi \leq T_{\min}/M \leq 6\sqrt{3}\pi$ & Upper bound at $Q=0$ \\
\hline
Kerr--Newman & $a^{2}+Q^{2} \leq M^{2}$ & $4\pi \leq T_{\min}/M \leq 6\sqrt{3}\pi$ & Upper bound at $a=Q=0$; \\
 &  &  & lower bound at $a=M,\ Q=0$ \\
\hline
Hayward & $0 \leq L \leq \dfrac{4M}{3\sqrt{3}}$ & $30.9 \lesssim T_{\min}/M \leq 6\sqrt{3}\pi$ & Upper bound at $L=0$ \\
\hline
Bardeen & $0 \leq g \leq \dfrac{4M}{3\sqrt{3}}$ & $28.4 \lesssim T_{\min}/M \leq 6\sqrt{3}\pi$ & Upper bound at $g=0$ \\
\hline
\end{tabular}
\end{table}

In Table~\ref{tab:summary}, we summarize the bounds on the minimum orbital
period for all five types of black holes investigated in this work. Several
regularities emerge from the comparison. First, for every black hole model,
the minimum orbital period is bounded by $4\pi M \leq T_{\min}
\leq 6\sqrt{3}\pi M$, regardless of whether the spacetime possesses
a central singularity. Second, the upper bound $6\sqrt{3}\pi M$ is
saturated whenever the black hole parameters vanish, i.e., in the
Schwarzschild limit ($Q = 0$, $L = 0$, or $g = 0$), and the normalized
minimum orbital period decreases monotonically as the charge, angular
momentum, or regularization parameter increases. Third, only rotation is
capable of pushing $T_{\min}$ to the lower bound $4\pi M$, which is
saturated exclusively by the maximally rotating extremal Kerr black hole.
For the static non-singular Hayward and Bardeen black holes, $T_{\min}/M$
remains well above the lower bound, with the smallest values $30.9$ and
$28.4$ attained at the critical parameter $4M/(3\sqrt{3})$, beyond which
the horizon disappears. These observations support the conjecture that the
bounds $4\pi M \leq T_{\min} \leq 6\sqrt{3}\pi M$ may be a
universal property of four-dimensional black holes with horizons.

\section{Conclusions}

In this work, we tried to search for the upper bound
on the minimum orbital period around black holes,
inspired by the known result that the minimum orbital period satisfies the lower bound
$T_{\min}\geq 4\pi M$, where $M$
is the black hole mass and $T_{\min}$ is the minimum orbital period.
Our analysis of Schwarzschild, Reissner-Nordstr\"{o}m and Kerr-Newman
revealed an upper bound on the minimum orbital period: $T_{\min}\leq 6\sqrt{3}\pi M$.
This bound exists regardless of different
values of charge or angular momentum. For Schwarzschild black holes,
the minimum orbital period occurs at $r=3M$, yielding $T_{\min}=6\sqrt{3}\pi M$.
In Reissner-Nordstr\"{o}m black holes, increasing the charge reduces the minimum
orbital period, so the upper bound remains unchanged. Similarly, in Kerr-Newman
black holes, increasing charge or angular momentum decreases the minimum
orbital period, so the upper bound is achieved in the non-rotating uncharged limit.
So the bounds $4\pi M \leq T_{\min} \leq 6\sqrt{3}\pi M$
hold for various types of singular black holes.

Besides singular black holes, we further investigated the minimum orbital periods
around non-singular Hayward and Bardeen black holes.
For the Hayward black hole, we found that increasing the quantum
gravity parameter $L$ leads to a decrease in $T_{\min}$,
while the upper bound remains unchanged and is saturated in the
Schwarzschild limit ($L=0$). Similarly, for the Bardeen black hole,
increasing the magnetic charge parameter $g$ also reduces $T_{\min}$,
with the upper bound again achieved in the Schwarzschild case ($g=0$).
Our results show that the bounds $4\pi M \leq T_{\min} \leq 6\sqrt{3}\pi M$ hold throughout the entire parameter
region in which the black hole horizon exists, with the lower bound never saturated
in the non-rotating Hayward and Bardeen spacetimes.
The fact that the smallest values of $T_{\min}$ ($\thickapprox$ 30.89 for Hayward and $\thickapprox$ 28.43 for Bardeen)
remain far above $4\pi M$, indicating that Hod's lower bound is never approached in these static spacetimes.

Notably, the upper bound $6\sqrt{3}\pi M$ is saturated by the standard Schwarzschild
black hole $(a = 0, Q = 0, L = 0, g = 0)$, while the lower bound $4\pi M$ is
saturated by the maximally rotating extremal Kerr black hole. All
other black holes lie between these two limits. If this pattern holds generally,
the bounds $4\pi M \leq T_{\min} \leq 6\sqrt{3}\pi M$ may constitute a universal theorem for
four-dimensional black holes, provided a horizon exists.
The bounds give a fundamental constraint on the
time period of particles orbiting a black hole, which affects
phenomena like accretion disk dynamics, and provides a perspective
for testing general relativity in extreme gravitational environments.
Previous studies have shown that orbital dynamics can be modified by
thermal and rotational effects \cite{RR1,RR2}.
It is therefore important to further test the universality of
these bounds in rotating or thermally-corrected black hole spacetimes.

\begin{acknowledgments}

This article is a merged and extended version of arXiv:2504.09061
and arXiv:2505.23000. The upper-bound analysis of Schwarzschild,
Reissner--Nordstr\"om and Kerr--Newman black holes was first obtained
by us in \cite{YPL}, whose content is fully incorporated here and
which is not intended for separate publication. A summary of the
bounds for all five black hole types is added.
This work was supported by the Shandong Provincial Natural Science Foundation of China under Grant
No. ZR2022MA074. This work was also supported by a grant from Qufu Normal University
of China under Grant No. xkjjc201906.

\end{acknowledgments}

\end{document}